\documentclass[10pt,onecolumn,a4paper]{article}

\newcommand{\affil}[1]{$^{\rm #1}$}
\usepackage[authoryear]{natbib}
\bibpunct{(}{)}{;}{a}{}{,}
\usepackage{graphicx}
\date{} 
\title{Interpretation of abundance ratios}
\author{{\it Francesca Matteucci\affil{A}, Cristina Chiappini \affil{B}}\\\\
\affil{A}\,Dipartimento di Astronomia, Universita' di Trieste, Via G.B. Tiepolo 11, 34127 Trieste, Italy\\
\affil{B}\,Osservatorio Astronomico di Trieste (INAF), Via G.B. Tiepolo 11, 
34127 Trieste, Italy\\}
\begin{document}
 \maketitle
\begin{minipage}{.9\textwidth}

{\bf Abstract}\\
In this paper we discuss abundance ratios and their relation to 
stellar nucleosynthesis and other parameters of chemical evolution models, 
reviewing and clarifying the correct use 
of the observed abundance ratios in several astrophysical
contests. 
In particular, we start from the well known fact that abundance ratios 
depend on
stellar yields, initial mass function and stellar lifetimes and 
we show,  by means of specific examples, that in some cases it
is not correct to infer constraints on the contributions from 
different SN types (Ia, II),  and particularly on different sets of yields,
in the absence of a complete chemical evolution 
model taking into account stellar lifetimes.
In spite of the fact that some of these results should be well 
known, we believe that it is useful to  discuss the meaning of abundance 
ratios 
in the light of severel recent claims based upon an incorrect interpretation
of observed abundance ratios.
In particular, the procedure, often used in the recent literature, of deriving 
directly conclusions 
about stellar nucleosynthesis, just by relating abundance ratios to 
yield ratios, implicitly assumes  the instantaneous recycling 
approximation (I.R.A.). 
This 
approximation is clearly not correct 
when one analyzes the contributions of SNIa relative to SNII  as functions 
of cosmic time.
In this paper we show that the uncertainty which arises from adopting this 
oversimplified 
procedure in a variety of astrophysical objects, such as 
elliptical galaxies, the intracluster medium and high redshift objects, does
not allow us to draw any firm conclusion, and that the differences between 
abundance ratios predicted by models with I.R.A. and models with detailed 
stellar lifetimes is of the same order as the differences between different 
sets 
of yields. On the other hand, if one is interested only in establishing the 
global
metal production (e.g. galaxies plus ICM) over the lifetime of the universe, 
then 
the adoption of simplified arguments can be justified.

\medskip{\bf Keywords:}

Nucleosynthesis, abundances, galactic evolution

\medskip
\end{minipage}

\section{Introduction}

Recently, a great deal of work has been done in measuring chemical abundances
not only in stars in the Galaxy but also in the intra-cluster medium (ICM) 
and 
in high redshift Damped-Lyman $\alpha$ (DLA) objects. Therefore, it is 
challenging 
for theorists to interpret the meaning of abundance ratios which are less 
affected than absolute abundances
by the assumptions of chemical evolution models, such as star 
formation rate, infall rate and outflow rate.

Abundance ratios, in fact, are largely independent of these parameters but
depend upon the stellar nucleosynthesis, the initial mass function (IMF) and
the stellar lifetimes. On the other hand, the variation of abundance ratios 
in time or 
as a function of metallicity depends on the star formation history as well.
Because of this,  it has been often proposed to use 
abundance ratios as a function of time or metallicity
as cosmic clocks, and in particular the ratios involving one element 
produced on short timescales and the other produced on large timescales, 
as for 
example O/Fe and N/O (Tinsley 1979). The bulk of iron and nitrogen are, in 
fact, produced on long timescales whereas oxygen is rapidly produced (on timescales of 
a few million 
years) by massive stars.
Iron production is due to type Ia supernovae (SNe) which are thought to originate from 
white dwarfs in binary systems and is restored over a range of timescales 
from $\sim 3 \cdot 10^{7}$ years to 15 Gyr and more. Nitrogen is mainly 
produced in stars with masses from 2 to 8 $M_{\odot}$ and therefore on 
timescales ranging from several tens of million years to some billion  
years.

Recent papers discussing the chemical abundances measured in the ICM 
derive 
constraints on 
the particular type Ia SN model just by comparing the measured abundance 
ratios with stellar production ratios in type II and Ia SNe.
In proceeding this way one implicitly assumes that abundance ratios can 
give direct information on the production ratios of the considered elements.

In this paper we show that the above assumption can be tolerated 
only if the abundance ratios concern elements produced on the same 
timescales but it is uncertain when applied to ratios such as O/Fe 
involving elements produced on quite different timescales.
Moreover, 
we point out that neglecting the stellar lifetimes
can be incorrect even for certain $\alpha$-elements such as S and Si when 
compared to O.
This is important in connection with the fact that a common procedure is to 
assume that $\alpha$-elements all evolve in the same way, and in particular 
to take oxygen as a proxy for Si and S when studying the evolution of DLA 
systems.

The paper is organized as follows:
in section 2 we discuss how the evolution of the abundance 
ratios depends on chemical evolution models. In sections 3, 4 and 5 we
discuss the chemical evolution of the solar neighbourhood, ICM and DLA
systems,  respectively. 
Finally, in section 6 we
draw some conclusions.

\section{Abundance ratios and chemical evolution of galaxies}
The Simple Model of chemical evolution  assumes
that the studied system evolves as a closed-box, that the IMF is constant 
in time, that the gas out of which the first stars form is primordial 
(i.e. Z=0)
and that the instantaneous recycling approximation (I.R.A.) holds 
(see Tinsley, 1980).
This approximation allows us to ignore the stellar lifetimes and therefore 
the delay with which some of the chemical elements are produced and 
restored into the ISM. In other words, all the elements are produced 
instantaneously.\par

The Simple Model predicts that the abundance of a generic metal 
$i$, $X_i$,
evolves as:
\begin{equation}
X_i=y_iln \mu^{-1}
\end{equation}
where $\mu = {M_{gas} \over M_{tot}}$ is the gas fraction in the system and 
$y_i$ is the ``yield'' per stellar generation:
\begin{equation}
y_i={1 \over (1-R)} \int_{m_{TO}}^{\infty}{mp_{im} \varphi(m) dm}
\end{equation}
which depends on the initial mass function $\varphi(m)$ (IMF) and the stellar yield $p_{im}$, namely the fraction of the stellar mass 
ejected as the newly created element $i$ by a star of mass $m$.
The mass $m_{TO}$ is the globular cluster``turn-off'' mass. 

The quantity 
$R$ is the so-called returned fraction:
\begin{equation}
R=\int_{m_{TO}}^{\infty}{(m-m_{rem}) \varphi(m) dm}
\end{equation}
with $m$ being the star mass and $m_{rem}$ the remnant mass.
It is worth noting that both $y_i$ and $R$ are fractions since they 
are both
divided by:
\begin{equation}
\int_{m_{TO}}^{\infty}{m \varphi(m)dm}=1
\end{equation}
which is the normalization condition for the IMF.
The IMF is generally expressed as:
\begin{equation}
\varphi(m)=A m^{-(1+x)}
\end{equation}
where the constant A is derived from eq. (4).

If the Simple Model holds, according to eq. (1) we can write that:
\begin{equation}
{X_i \over X_j} = {y_{i} \over y_{j}}
\end{equation}
Therefore, in this case the ratio of two abundances is directly the 
ratio of the two ``yields'' defined in equation (2).
The same is true if instead of the Simple Model we consider models with 
inflows/outflows but always with I.R.A. approximation.

For example, in the presence of outflow and under the assumption 
that the outflow rate is simply proportional to the star formation rate 
multiplied by $(1-R)$ through a constant $\lambda$, one can find the 
following solution for metals
(see Matteucci \& Chiosi, 1983; Matteucci 2001):

\begin{equation}
X_i={y_i \over (1- \lambda)}ln[(1 + \lambda) \mu^{-1} - \lambda]
\end{equation}
where $\lambda$ is a constant larger than zero.
Again, the abundance ratio
of the two elements coincides with the ratio of their yields.

A similar situation occurs for an infall model with I.R.A. 
when there is infall of primordial material without metals ($(X_{i})_{inf}=0$) and the infall 
rate is proportional to the star formation rate multiplied by $(1-R)$ 
through a positive constant 
$\Lambda \ne 1$ (Matteucci \&
Chiosi, 1983; Matteucci 2001):
\begin{equation}
X_i={y_i \over \Lambda}[1-(\Lambda-(\Lambda-1) \mu^{-1})^{-{\Lambda \over (1- \Lambda)}}]
\end{equation}
If $\Lambda=1$ the solution, always for metals, is:

\begin{equation}
X_{i}=y_i[1-e^{-(\mu^{-1}-1)}] 
\end{equation}
which is the well-known solution for the {\it extreme infall}  
(Larson, 1972),  where the amount of gas in the system remains constant 
in time. 

Finally, in the case of infall and outflow operating
at the same time the solution is:
\begin{equation}
X_i={y_i  \over \Lambda}\{1-[(\Lambda - \lambda)-(\Lambda - \lambda - 1) \mu^{-1}]
^{{\Lambda \over \Lambda - \lambda -1}}\},
\end{equation}
for a primordial infalling gas ($(X_{i})_{inf}=0$) as shown in Matteucci (2001).

In all of these cases (7,8,9,10) it is evident that the equation (6) holds, but 
this is not true anymore if one relaxes  I.R.A.,  as is shown in 
Figure 1 where we plot the predicted O/Fe abundance ratio (abundances by mass) as a function of time,  as predicted
by a 
detailed model of the Milky Way (Chiappini et al. 2003). As one can see, 
in fact, the [O/Fe] ratio is strongly varying with time and this is due to 
the delayed Fe production relative to the O production.

\begin{figure}[h]
\begin{center}
\includegraphics[scale=0.5, angle=0]{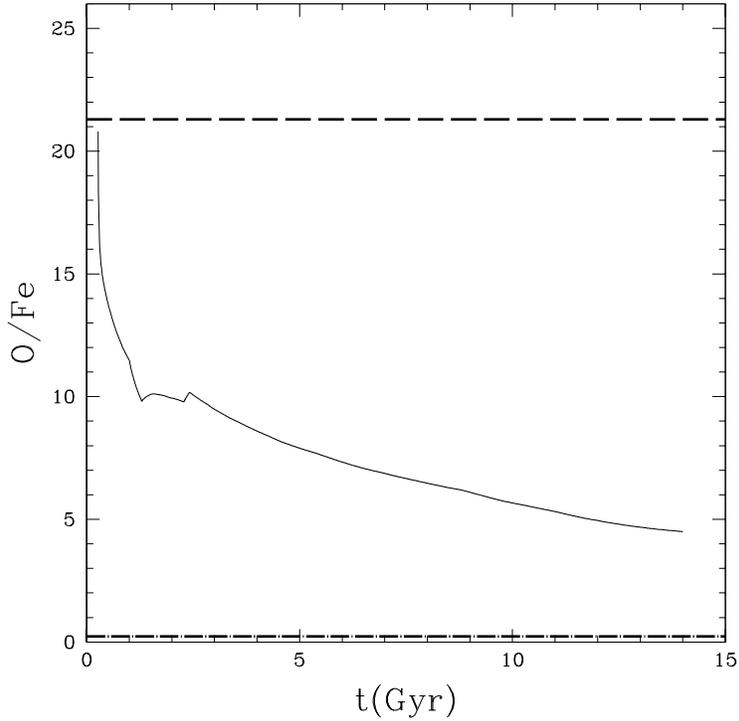}
\caption{Predicted 
O/Fe abundance ratio (abundance by mass) as a  function of time, as predicted
by the time-delay model of 
chemical evolution for the solar vicinity (Chiappini et al. 2003).
Also shown are the (O/Fe)$_{SNII}$ (dashed line) and 
(O/Fe)$_{SNIa}$ (dash-dotted line) production
ratios as given in Nomoto et al. (1997) both for a type Ia SN and a 
20$M_{\odot}$ star taken as representative 
of an average massive star exploding as type II SN.} 
\label{fig 1}
\end{center}
\end{figure}

This delay allows us to interpret the [O/Fe] vs. [Fe/H] observed for solar 
neighbourhood stars (Greggio \& Renzini 1983; Matteucci \& Greggio 1986). In the same figure we show the
constant  O/Fe production ratio expected from  a typical SNII and a 
typical SN Ia (dash and dot-dashed lines, respectively).
In particular, we took the production ratio of a 20 $M_{\odot}$ star, as representative of a typical massive star, from
Nomoto et al. (1997).  

On the other hand, for the ratios between elements produced on similar 
timescales, such as 
the ratios between different $\alpha$-elements which are mainly originating 
from massive 
short-lived stars, one can reasonably assume that they are almost constant 
during the whole galactic lifetime.

\begin{figure}[h]
\begin{center}
\includegraphics[scale=0.5, angle=0]{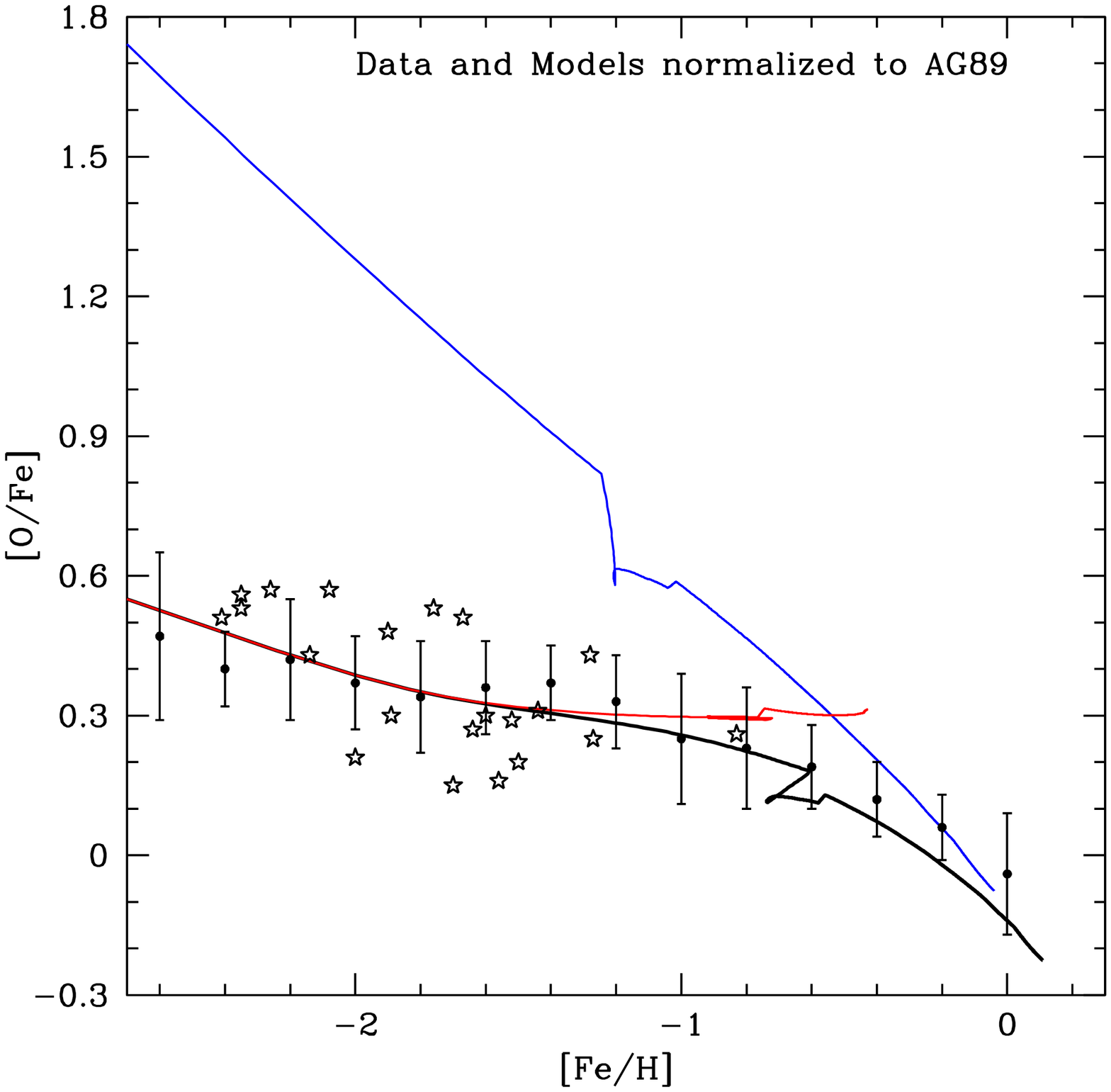}
\caption{Predicted [O/Fe] vs. [Fe/H] from the model of Chiappini et al. 
(2003) (thick continuous line), compared with the data from Mel\'endez \& Barbuy 
(2002). The strongly decreasing line represents the prediction of the model when 
the Fe production from type II SNe is suppressed, whereas the almost constant
line refers to the 
predictions when the Fe production from type Ia SNe is suppressed.}
\label{fig 2}
\end{center}
\end{figure}

Therefore, the assumption that abundance 
ratios reflect stellar yields is true in a first approximation only for 
elements produced on the same timescales, such as $\alpha$-elements, although 
also in this case there are exceptions, as we will show in section 5.

\section{The roles of SNe in the [O/Fe] versus [Fe/H] plot for the Milky Way}
In Figure 2 we show the [O/Fe] vs. [Fe/H] relation
as predicted by the best model for the solar neighbourhood of 
Chiappini et al. (2003) (model 4 with yields of Nomoto et al. 1997). 
The model reproduces very well the most recent and accurate 
measurements in stars of the solar neighbourhood.
The model is a two-infall model where the halo-thick disk forms on a 
relatively short timescale ($\sim 1-2$ Gyr) whereas the disk forms on 
much longer timescales and in an ``inside-out'' fashion.
The best model takes into account detailed nucleosynthesis from type II and 
Ia SNe, and predicts that type Ia SNe produce $\sim70\%$ of the present 
time Fe abundance whereas the remaining
$30\%$ comes from type II SNe. This prediction is a direct consequence 
of the assumed stellar yields and IMF. The yields of O and Fe from massive 
stars 
and the yields of Fe from type Ia SNe are from 
Nomoto et al. (1997), their model W7, while for the IMF we assume that of 
Scalo (1986).
On the same figure we show one model where we arbitrarily assumed that all 
the Fe would arise from type II SNe and one model where all the Fe 
originates 
from type Ia SNe. All the models  and the data are normalized to Anders and 
Grevesse (1989) solar
values.
As one can see, neither of the two models would fit the observed pattern
since Fe originating only from type II SNe would produce a flat 
[O/Fe] over all the 
[Fe/H] range, as expected. On the other hand, if all the Fe were to originate 
from type Ia SNe then the [O/Fe] would be linearly decreasing with 
increasing metallicity without showing 
any plateau at low metallicities. Clearly, to explain the observed pattern we 
need both SN types and,  to do that, we need to  relax the I.R.A. 
approximation.
For this reason, this kind of model is often referred to as  
``time-delay model''. 
In fact, under I.R.A. we would predict 
always a constant [O/Fe] for the whole metallicity range. 
Therefore, in conclusion, ignoring the effect of stellar timescales produces 
an incorrect
interpretation of data of any kind. 

Intuitively, one may think that after a very long 
time since the end of star formation, when the gas content 
tends to zero,
the abundance ratios will tend to the 
ratios
of their yields per stellar generation (eq.2). This is true, in principle, if the global metal production is considered 
(namely the metals in stars, in gas inside and outside galaxies), but it fails if only the metals in the individual components (e.g. in the gas)
are taken into account.
In fact, as shown by Prantzos \& Aubert (1995), just when the gas in a system tends to zero then the effects of relaxing I.R.A. are the strongest.
In particular, the final amount of gas in the system is influenced by relaxing I.R.A., since low mass stars restore 
their external envelopes at the present time
and these envelopes contain the abundance patterns of the gas in the early 
stages of galactic evolution.
The effect of this delayed gas return is negligible
in galaxies with 
ongoing star formation and therefore possessing a relatively high fraction of gas, whereas it is important in objects which have stopped 
forming stars 
several Gyrs ago, such as ellipticals, as we will show in 
the next paragraph.

\section{Abundances in the ICM}

Several papers have already appeared in the literature trying to interpret
the measured abundance ratios in the ICM simply in terms of stellar 
yields, with 
the consequence of imposing constraints on different nucleosynthesis models
for type Ia SNe as well as on the different proportions of Fe produced by 
type Ia and II SNe 
(see for example Gibson, Lowenstein \& Mushotzky, 1997, Finoguenov et al. 2002, Gastaldello \& Molendi 2002).
For instance, the empirical method adopted in the observational papers dealing with ICM abundance ratios
can be summarized as follows:
the observed ratios of several elements to iron are used in order to find
the best fit to a function defined as:
\begin{equation}
({X/Fe \over X_{\odot}/Fe_{\odot}})_{observed}=f({X/Fe \over X_{\odot}/Fe_{\odot}})_{SNIa} + (1-f)({X/Fe \over X_{\odot}/Fe_{\odot}})_{SNII},
\end{equation}
where the ratios labelled $SNIa$ and $SNII$  are the ratios of the stellar 
yields expected from SNIa and II relative to the solar abundance ratios, 
respectively. The quantity $f$ represents the unknown fraction of Fe 
produced by SNIa relative to the total Fe produced by type Ia plus 
type II SNe.
In principle, equation 11 is valid if one is interested in the global metal 
production (stars, gas inside and outside galaxies),
but it provides a poor approximation if one studies the abundances in the 
different components individually, such as the abundances in the ICM
and especially their evolution with cosmic time.
One reason for this resides in the fact that I.R.A. is a bad approximation 
for studying the evolution of the Fe abundance.

\begin{figure}[h]
\begin{center}
\includegraphics[scale=0.5, angle=0]{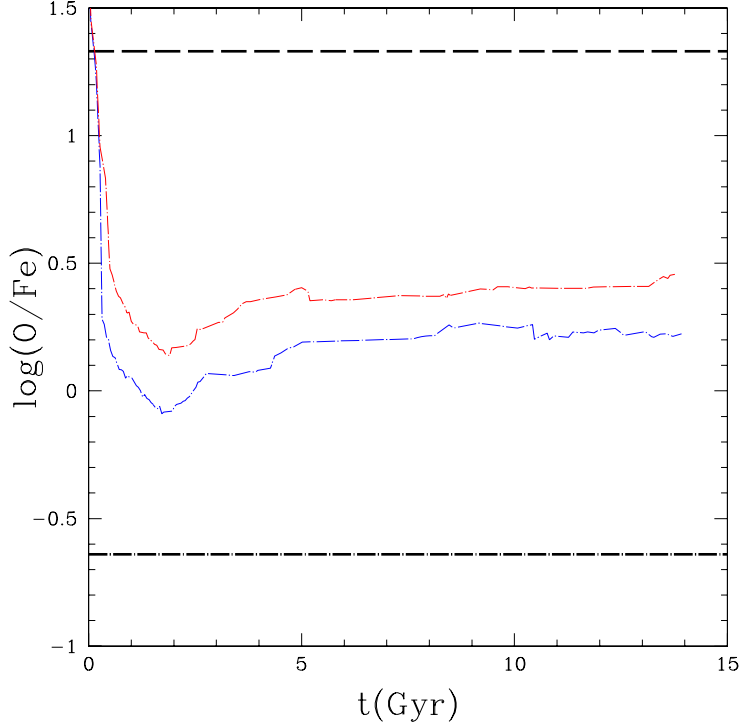}
\caption{Predicted log (O/Fe) versus time in the ISM of elliptical galaxies
before and after the occurrence of galactic winds which start at early 
times ($\le 0.3$ Gyr) and continue until the present time.
The model predictions are from Pipino et al.(2002), their Model I, 
inverse wind case, with Salpeter IMF.
The galaxies have masses of  $10^{10}$ M$_{\odot}$ (lower curve) and 
of $10^{11}$ M$_{\odot}$
(upper curve). The abundances are by mass.
The constant lines  represents the $(O/Fe)_{yields}$ ratio for
massive stars (dashed line) and for type Ia SNe (dashed-dotted)
from Nomoto et al. (1997) (W7 model for type Ia SNe), 
as in Fig.1.}\label{fig 3}
\end{center}
\end{figure}

Moreover, in the approach described by eq. 11, what is completely ignored 
is the mechanism of chemical enrichment
of the ICM from galaxies which strongly affects the ICM abundance ratios.
Several steps are necessary to compute the ICM chemical abundances. The first one is to compute the chemical evolution of elliptical
galaxies.
Elliptical galaxies are, in fact, believed to be the main contributors to the ICM
enrichment (e.g. Matteucci \& Vettolani, 1988; Arnaud et al. 1992; Matteucci \& Gibson, 1995; 
Renzini 1997; Pipino et al. 2002; Chiosi, 2000; Moretti et al. 2003).
In Figure 3 we show the evolution of the ratio of a typical 
$\alpha$-element such as O relative to Fe as a function of time, 
as predicted for the gas in  elliptical galaxies of luminous 
mass $\sim 10^{11}M_{\odot}$ and $\sim 10^{10}M_{\odot}$. 
In the same figure is shown the constant O/Fe ratio derived from the ratio 
of the yields (production ratios) of oxygen and iron 
(for SNII - dashed line and 
SNIa - dot-dashed line).
As one can clearly see from the figure, the O/Fe ratios in the gas in 
ellipticals,  after the star formation stops,  are not constant 
but they first decrease, as expected because of the occurrence of more and more 
SNIa, then increase and finally reach a plateau 
at late times. 
This latter effect, often forgotten, is
due to the relaxation of I.R.A. and is produced by 
the low mass stars formed out of gas enriched in oxygen relative to iron 
at early times, as already mentioned. 
These stars, in fact, eject through the 
planetary nebula phase their pristine gas with high O/Fe ratio and
this gas can be later on restored into the ICM 
by means of galactic winds and/or ram pressure stripping.
This is the case of figure 3 since the models presented here predict a 
continuous galactic wind
since early times. In particular, the galactic winds start before 0.3 Gyr 
in both galaxies, thus most
of the Fe produced by type Ia SNe is ejected into the ICM.
It is worth noting that the final O/Fe ratios contributed by galaxies of 
different luminous masses 
($10^{10}$, $10^{11}M_{\odot}$) are different and the difference  
depends on the duration of the star formation period in each galaxy. This 
occurs because
the degree of $\alpha$-enhancement depends on the timescale of the
Fe released by type Ia SNe relative to the timescale of star formation 
(see Matteucci \& Recchi 2001). 
Good models of chemical evolution for elliptical galaxies predict that the 
star formation
timescale was shorter in more massive objects, in order to reproduce the 
observed increase
of the [$\alpha$/Fe] ratio in the dominant stellar population as a function 
of galactic mass
(Matteucci 1994; Pipino \& al. 2002; Pipino \& Matteucci 2003, Romano et al. 
2003).
This implies that galactic winds should occur first in massive objects, 
thus interrupting the
star formation process.
As a consequence of this,
the final log(O/Fe) produced by a $10^{10}M_{\odot}$ (luminous mass)
galaxy is $\sim 0.2$ whereas that produced by a $10^{11}M_{\odot}$ galaxy 
is $\sim 0.5$. The difference, $\Delta log(O/Fe) \sim 0.3$, is comparable to
the differences among production ratios obtained by adopting different sets 
of yields. 
For example, if we take the SNII oxygen yield and the yields of Fe 
predicted by the different type Ia models of Iwamoto et al. (1999) 
we find that the maximum difference between production ratios is 
$\Delta log(O/Fe) \sim 0.7$ 
(between model WDD3 and model CDD1 which produce the highest and lowest 
Fe in type Ia SNe, respectively).

In realistic models for the chemical enrichment of the ICM (e.g. Matteucci \& Vettolani 1988, Matteucci 
\& Gibson 1995, Pipino et al. 2002) one then should
integrate the contribution of single cluster galaxies to the O and Fe enrichment in 
the ICM by means of the cluster luminosity function.
The total masses of O and Fe which are ejected into the ICM crucially depend on the assumptions
about the duration of galactic winds and/or ram pressure stripping, in other words about how much mass is lost from the galaxies
(generally the mass at the break of the luminosity function dominates the ICM enrichment).
In particular, in Matteucci \& Vettolani (1988) and Pipino et al. (2002) it was found that the Fe mass observed
in galaxy clusters, either poor or rich, can be reproduced  with a Salpeter (1955) IMF, only if all the Fe produced
by type Ia SNe, after the star formation has stopped, is sooner or later restored into the ICM.
On the other hand, a top-heavy or a variable IMF (Matteucci \& Gibson, 1995, Chiosi, 2000; Moretti et al. 2003)
can reproduce all the observed ICM Fe with the only contribution of early galactic winds.

A top-heavy IMF rather than a Salpeter -like IMF increases the
[$\alpha$/Fe] ratio predicted for the dominant stellar population in 
ellipticals,
but it does not prohibit obtaining an almost solar
[$\alpha$/Fe] ratio in the ICM at the present time, if one assumes that 
all the Fe produced by
type Ia SNe, after  star formation ceases, sooner or later will be restored into the ICM; then, however, 
the total Fe mass restored into the ICM
is overestimated. In this case, one is forced to assume that most of the Fe produced by type Ia SNe 
has been retained by the galactic potential well and has not been mixed with the ICM.
As a consequence of this, one predicts an overabundance of oxygen relative to iron 
([$\alpha$/Fe] $>0$) in the local ICM. 

Therefore, it is clear that because of all the effects mentioned above the 
particular effect of the stellar yields on the O/Fe ratio in the ICM 
is quite difficult to extract from the measured abundance ratios.
As a consequence, no firm conclusion about the different nucleosynthesis 
models for type Ia SNe can be drawn on the basis of the simple 
assumption that abundance ratios are indicative of the yield ratios.

What is instead a robust conclusion is that if the observed [$\alpha$/Fe]
ratio in the ICM is solar or slightly undersolar (although the measured abundance ratios in clusters are still uncertain
and do not allow us to draw firm conclusions, 
see Loewenstein 2004 
for a recent review on the subject), one can safely conclude {\bf that the contribution of type 
Ia SNe to the Fe enrichment, relative to the Fe enrichment from type II SNe, has been the same as in the solar vicinity.}
In fact, if the [$\alpha$/Fe] enhancement observed in the Galactic halo 
stars is interpreted as being due to the prompt enrichment by type II SNe and that 
the almost solar [$\alpha$/Fe] seen in the Galactic disk stars is due to the 
addition of Fe from type Ia SNe (see Figure 2), then the same reasoning should hold for 
the ICM (Renzini 1997), irrespective of the differences among 
different sets of yields. This is at variance with the conclusion of 
Gibson et al. (1997) who stated that, owing to the uncertainties still 
present in the stellar yields, no firm conclusions on the role of type Ia 
and II SNe in the ICM enrichment can be drawn. 
In this case, in fact, only those yields  should be used which 
best reproduce the [$\alpha$/Fe] vs. [Fe/H] in the solar neighbourhood, 
given the supposed universality of nucleosynthesis. 

\section{Damped-Lyman $\alpha$ Systems}

Another important application of the abundance ratios is to infer the 
nature and the age of DLAs, namely the absorbers of quasar light observed 
at high redshift. In particular, these objects possess a high neutral gas 
content ($N_{HI} \ge 2 \cdot 10^{20} cm^{-2}$) and metal abundances ranging 
from  $\sim 1/100$ up to $\sim 1/3$ of the solar value 
(from [Fe/H] $>-$2.0 dex).
Since different histories of star formation in different galaxies produce 
different [$\alpha$/Fe] vs. [Fe/H] relations, as first pointed out by 
Matteucci (1991) and as shown in
Figure 4, 
one can try to infer the nature of DLA systems by comparing the abundance 
patterns with galactic chemical evolution models (see Calura et al. 2003).
In Figure 4 we show the [$\alpha$/Fe] vs. [Fe/H] (where $\alpha$ indicates O+Mg) relations predicted for 
different histories of star formation: for a system with intense star 
formation rate, such as the galactic Bulge, the predicted [O/Fe]  stays 
constant for 
a longer [Fe/H] interval than in  the solar vicinity with milder star 
formation 
and in an 
irregular magellanic system with even milder star formation.

\begin{figure}[h]
\begin{center}
\includegraphics[scale=0.5, angle=0]{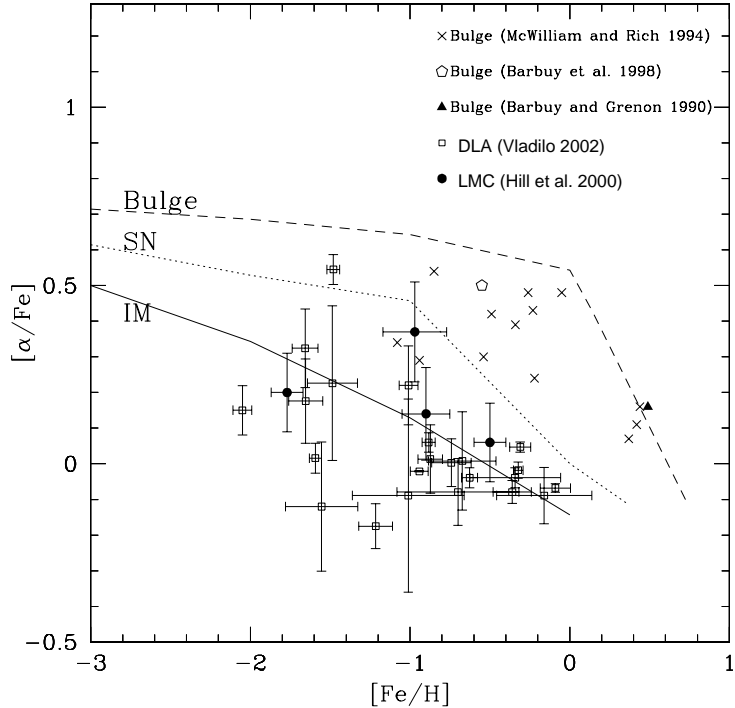}
\caption{Predicted [$\alpha$/Fe] versus [Fe/H] for different histories of 
star formation. In particular, the upper curve represents the galactic Bulge, 
the central one the solar vicinity and the lower one an irregular 
magellanic galaxy. The data refers to the Bulge stars and to DLA systems,
as explained in the text.}\label{fig  4}
\end{center}
\end{figure}

\begin{figure}[h]
\begin{center}
\includegraphics[scale=0.5, angle=0]{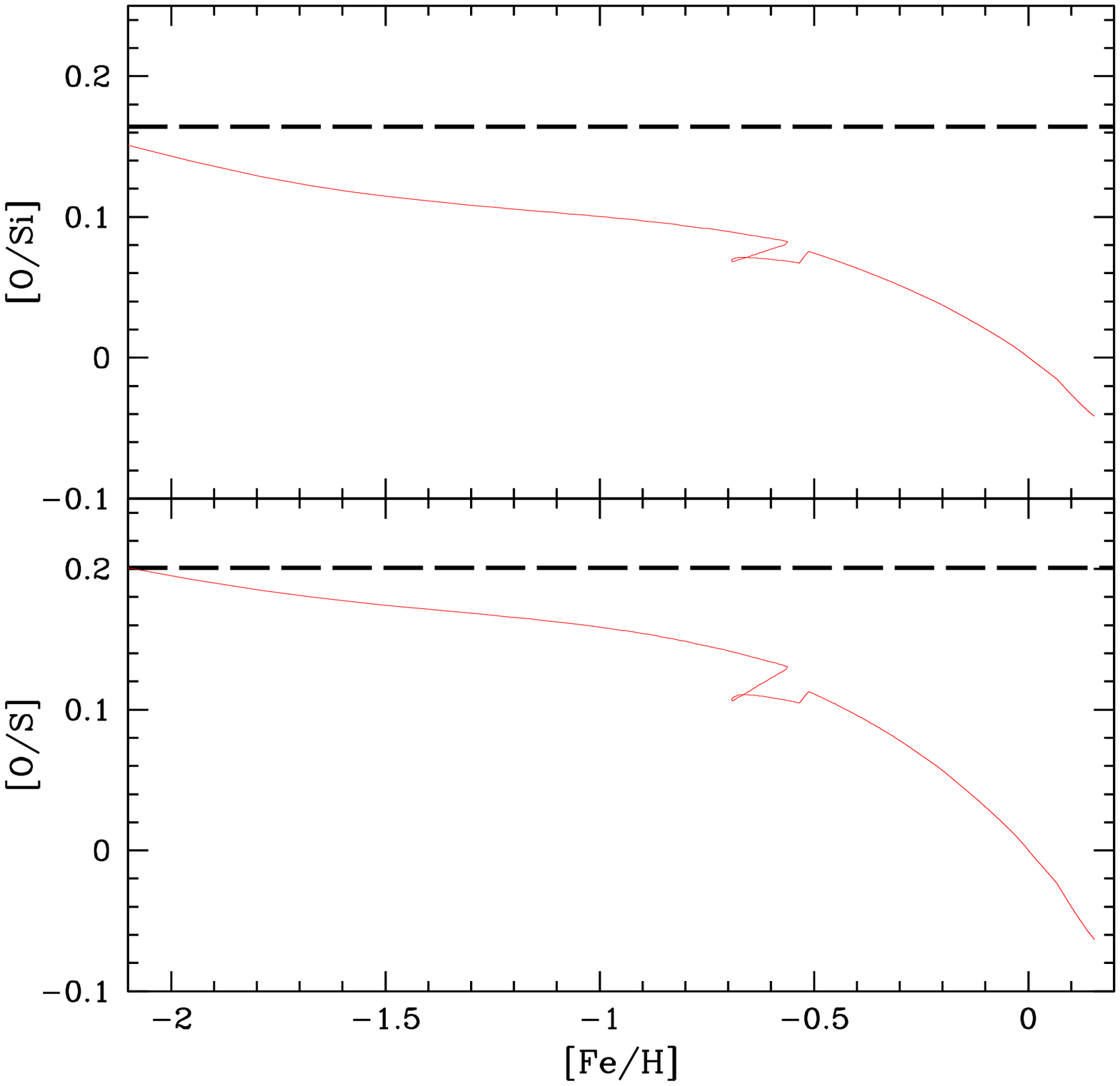}
\caption{Predicted [O/S] and [O/Si] as functions of [Fe/H]  
by the model of Chiappini et al. (2003) for
the solar vicinity. 
The constant lines represents the $(O/Si)_{yields}$ and $(O/S)_{yields}$ 
ratios for
massive stars  
taken from Nomoto et al. (1997).}\label{fig 5}
\end{center}
\end{figure}

The different behaviours of the [$\alpha$/Fe] ratio can easily be understood on the 
basis of the time-delay model described in section 3.
A comparison with observed abundance ratios in DLA systems indicate that these 
objects are likely to be irregular galaxies rather than spheroids such as 
bulges and  elliptical galaxies. Therefore the [$\alpha$/Fe] ratios versus [Fe/H]
represent a unique diagnostic for revealing the nature of high redshift objects.

Often in the literature (Pettini et al., 2002;  Prochaska et al. 2002; Centurion et al. 2003), when dealing with DLA systems, S and 
Si are used as proxies for O, because they are all $\alpha$-elements.
However, also among $\alpha$-elements there are differences, in particular 
between O on the one hand and S and Si on the other.
These latter elements, in fact, according to the majority of the available 
stellar yields,  are produced in a non-negligible way by 
SNe Ia whereas oxygen is entirely produced by massive stars. This 
difference is reflected in the predicted O/S and O/Si ratios, as shown 
in Figure 5.
There it is clear that the O/S and O/Si ratios predicted
for the solar neighbourhood are not constant as functions of [Fe/H] and therefore time, as 
one would expect if S and Si were proxies for oxygen.
The variation of the predicted O/S and O/Si abundance ratios
are partly due to the mass dependence of their production ratios in 
massive stars and partly to the fact that S and Si are produced in type Ia SNe more than is oxygen. 
We recall that the very first SNe Ia, in the framework of the 
progenitor model adopted here (single degenerate scenario, see 
Matteucci \& Recchi 2001), occur already after $\sim 30-40$ million years since the beginning of star formation.
The deviation of the predicted O/S and O/Si abundance ratios from the corresponding production ratios in a typical
massive star, as shown in Figure 5, is non-negligible especially in the observed range of DLA
systems.

Therefore, there is a danger in interpreting abundance ratios relative to 
Si and S as if they were ratios relative to oxygen.

\section{Conclusions}

In this paper we discussed the correct use of abundance ratios
in order to infer valuable constraints on the stellar nucleosynthesis 
and the star formation history in galaxies.
In particular, we have pointed out that:
\begin{itemize}
\item
abundance ratios depend not only 
on stellar yields and IMF but also on the timescales of production of 
the various elements. 
This means that, when relaxing the hypothesis of instantaneous recycling (I.R.A.),
the abundance ratios are not good indicators of yield ratios, as is often 
assumed in the current literature.
\item
Observed abundance ratios at the present time in galaxies with a small gas content are affected 
by the often ignored effect 
of late gas pollution due to low mass stars restoring, at the present time, 
the H-rich and $\alpha$/Fe
enhanced gas out of which they formed at early times.
We have calculated the specific case of elliptical galaxies which are 
believed to have stopped forming stars in a substantial way several 
Gyr ago, and have shown that 
the predicted present time abundance ratios (e.g. $\alpha$/Fe)
do not reflect the yield ratios (we adopted the same yields and IMF) 
but vary according to the initial luminous galactic mass which  
influences the history of star formation and galactic winds in each object.
In particular, the differences between the abundance ratios of ellipticals of different mass 
are comparable to the differences among different 
sets of yields.
As a consequence, it is a risky procedure to impose constraints upon 
different 
sets of yields for SN II and Ia directly from the observed abundance ratios, 
especially from the abundance ratios measured in the ICM.
A correct interpretation of these abundances requires a detailed galactic 
model able to follow the evolution of the absolute abundances in time 
plus specific assumptions about the
amount of gas ejected into the ICM by ellipticals.
The direct comparison of abundance ratios with the yields per stellar 
generation is, in principle, a valid procedure if one is interested 
only in computing the global metal production,
including gas and stars in galaxies plus the ICM, over the lifetime of 
the universe (see Calura \& Matteucci, 2004 for a detailed calculation 
of the global metal production in galaxies plus ICM/IGM).

\item As a consequence of the previous point, the observed [$\alpha$/Fe] 
ratio in the ICM depends not only on the yield
ratios but also on the star formation timescales in ellipticals. These in turn are related to 
the time of occurrence of a galactic wind, and to the question whether all the Fe
produced by type Ia SNe, after star formation has stopped, is soon or later restored into the ICM or is partly retained by the
galactic potential well.
Unfortunately, the observed abundance ratios in the ICM are still too uncertain to draw firm conclusions such as
suggesting particular IMFs for galaxy clusters and in any case it would be better to consider the global metal production including galaxies plus ICM
(see Portinari et al. 2004 for a discussion of these issues).
\item
Also the ratios of different $\alpha$-elements 
especially O on one side and Si and S on the other, are not constant 
in time, as expected in a first approximation, partly because O/Si 
and O/S production ratios are not constant in massive stars, but mainly because
Si and S are produced in a non negligible way also by long living type Ia SNe.
For this reason, Si and S cannot be used safely as proxies for O, as is often 
assumed in interpreting data for DLA systems, since the variation
of the [O/Si] and [O/S] ratios over the metallicity range typical of DLAs can be as high as $\sim$ 0.3  and $\sim$ 0.2 dex, respectively.
\end{itemize}

\section*{Acknowledgments} 
We are indebted with S. Recchi, F. Calura and A. Pipino for their invaluable help.
We also thank an anonymous referee for her/his careful reading and very 
useful suggestions.
We acknowledge financial support from the COFIN2003 project (MIUR) 
Prot. N.2003028039.

\end{document}